\documentclass[aps,showpacs,preprint]{revtex4}
\usepackage{amsmath}
\usepackage{epsfig}

\begin{document}

\title{Non-relativistic calculation of longitudinal 
$(e,e')$ response functions of trinucleons
at intermediate momentum transfers}

\title{Extension of non relativistic calculations of (e,e') longitudinal response functions at
higher momentum. How to minimize the frame dependence? }

\title{How to minimize the frame dependence of non relativistic
calculations of (e,e') quasi elastic response functions: a test on $^3$He}

\title{Improved $(e,e')$ response functions at intermediate momentum transfers:
the $^3$He case}

\author{Victor D. Efros$^{1}$\footnote{ on leave of absence from 
Russian Research Centre 
"Kurchatov Institute",  123182 Moscow,  Russia},
  Winfried Leidemann$^{1}$, 
  Giuseppina Orlandini$^{1}$
  and Edward L. Tomusiak$^{2}$ 
  }

\affiliation{
  $^{1}$Dipartimento di Fisica, Universit\`a di Trento, and
  Istituto Nazionale di Fisica Nucleare, Gruppo Collegato di Trento,
  I-38050 Povo, Italy \\
  $^{2}$Department of Physics and Astronomy,
  University of Victoria, Victoria, BC V8P 1A1, Canada\\
}

\date{\today}

\begin{abstract}
A possibility of extending the applicability range of non-relativistic
calculations of electronuclear response functions in the quasielasic peak 
region is studied. We show that adopting a particular 
model for determining the kinematical inputs of the non-relativistic
calculations can extend this range considerably, almost eliminating 
the reference frame dependence of the results.
We also show that there exists one reference frame,
where essentially the same result can be obtained with no need of 
adopting the particular kinematical model.
The calculation is carried out with the
Argonne V18 potential and the Urbana IX three-nucleon interaction. 
A comparison of these improved calculations with experimental data shows 
a very good agreement for the 
quasielastic peak positions at $q=500,$ 600,
700 MeV/c and for the peak heights at the two lower $q$--values, while for 
the peak height at $q=700$ MeV/c one finds differences of about 20\%.
\end{abstract}

\bigskip


\maketitle

\renewcommand {\thefootnote}{\fnsymbol{footnote}}
In Ref. \cite{elot04} we have studied the longitudinal response functions for 
electron scattering from three--nucleon systems in the momentum transfer 
range between 250 and 500 MeV/c. A non-relativistic ( n.r.) formulation of 
the nuclear three--body problem has been adopted and the full dynamics has been taken into account both in the initial and final state. A related study has 
been recently presented in Ref. \cite{bochum04}. In order to check the validity 
of our n.r. calculation we have checked in \cite{elot04}, among other
issues, also the reference frame dependence and found that it is not negligible 
for momentum transfers $q \ge 400$~MeV/c. A frame dependence of a similar type had already been 
observed in deuteron electrodisintegration \cite{BA1,BA2,BA3}. In Ref. \cite{elot04}
the hadronic current was evaluated in the Breit frame and the results were compared
with experimental data.  In the present work we 
reconsider the frame dependence and present results up to $q=700$~MeV/c.

It is clear that as $q$ increases the results of purely n.r. 
calculations must become increasingly questionable. One manifestation of
the importance of relativity is the frame dependence that occurs in such
n.r. calculations at high $q$. Of course use of any frame in a genuine relativistic
calculation must lead to the same laboratory (LAB) frame result. We will show that certain
frames in a n.r. calculation may tend to minimize the error due to lack
of a proper relativistic calculation. We also
suggest a procedure to reduce the frame dependence in the
quasi-elastic peak region.

In the one photon exchange approximation the inclusive electron scattering
cross section in the LAB frame is given by 
\begin{eqnarray} \label{X}
\frac{d^2\sigma}{d\Omega\,d\omega}\ =\ \sigma_M\ \bigg[\frac{(q^2-\omega^2)^2}{q^4}\,
R_L(q,\omega)+ \left(\frac{(q^2-\omega^2)}{2q^2}+\tan^2\frac{\theta}{2}\right)  
\, R_T(q,\omega)\bigg] \,,
\end{eqnarray}
where $R_L$ and $R_T$   
are the LAB longitudinal and transverse response 
functions, respectively.  
The LAB frame electron variables are denoted by $\omega$ 
(energy transfer), $q$ (momentum transfer), and $\theta$ (scattering angle).

In addition to $R_L$ one may define related responses $R_L^{fr}$ expressed in terms
of quantities pertaining to reference frames obtained via boosting 
the LAB frame along {\bf q}. In general nuclear states are products of
internal  and center of mass momentum substates. In the n.r.
approximation after integrating over the center of mass momentum one has
\begin{equation}\label{rfr}
R_L^{fr}=\sum\!\!\!\!\!\!\!\int df
\left|\langle\psi_i|\sum_j\ \rho_j({\bf q}^{fr},\omega^{fr})|\psi_f\rangle\right|^2
\delta\left(E_f^{fr}-E_i^{fr}-\omega^{fr}\right)\,.
\end{equation}
Here $q^{fr}$ and $\omega^{fr}$ are the momentum and energy transfer in a new reference frame, 
the internal substates are indicated with $\psi_i$ and $\psi_f$, and
$\rho_j({\bf q}^{fr},\omega^{fr})$ are the internal single--nucleon charge operators as defined in 
\cite{elot04} (the energy dependence is due to the inclusion of the nucleon form factors). 

The  summation-integration symbol 
denotes the usual sum/integration over final state variables in addition to 
averaging over the initial state magnetic quantum numbers. In the relativistic
case we have the same formula, but with the substates $\psi_i$ and $\psi_f$
depending, respectively, on the total momenta ${\bf P}_i^{fr}$ and 
${\bf P}_f^{fr}={\bf P}_i^{fr}+{\bf q}^{fr}$ of the initial and final states
in a given reference frame. Thus $\psi_i$ and $\psi_f$ are frame dependent
in the relativistic case. We disregard this frame dependence
in our calculations and we do not consider the boost corrections of 
the states.

Energy conservation is explicit in the
argument of the $\delta$--function where $E_f^{fr}$ and $E_i^{fr}$ denote the total initial and 
final energies  and can be expressed with relativistic or n.r. kinematics (both cases will be 
considered in the following). In the n.r. case the center of mass and internal energies can be 
separated so that:
\begin{eqnarray}\label{delta}
\delta\left(E_f^{fr}-E_i^{fr}-\omega^{fr}\right)\, & \approx &\,\delta\left(e_f^{fr} + 
\left(P_f^{fr}\right)^2/(2M_T)
\ -\ e_i^{fr} - \left(P_i^{fr}\right)^2/(2M_T) - \omega^{fr}\right)\\
& \equiv & \delta\left(e_f^{fr} - e_f^{nr}(q^{fr}, \omega^{fr})\right)
\end{eqnarray}
where $e_f^{fr}$, $e_i^{fr}$,  are intrinsic energies  of the final and initial states.

The response $R_L$ can be expressed in terms of $R_L^{fr}$ with the help of
the relationship 
\begin{equation}\label{facs}
R_L(q,\omega) = \frac{q^2}{(q^{fr})^2}{\frac {E_i^{fr}} {M_T}}
 R_L^{fr}(q_{fr},\omega_{fr})\,.
\end{equation}
The origin of the factor $q^2/(q^{fr})^2$ is shown in Ref. \cite{BA1}, Eqs. (2.13), (2.14) (see also e.g. Refs. \cite{BA2,ALT,DG}). The factor $E_i^{fr}/M_T$ arises since
we adopt the usual normalization of the target state to
unity instead of its covariant normalization.  (In [1] this factor
was not included). 

We will use relation  (\ref{facs}) to get 
the LAB response from  calculations 
referring to frames different from the LAB frame and study the frame dependence of
n.r. calculations.

In addition to the LAB frame we consider three other frames.  One is
the so--called anti--lab (AL) frame, where 
the total momentum in the final state is zero. Thus the target nucleus 
has a momentum $-{\bf q}_{AL}$. If one neglects the internal motion of the
nucleons inside the nucleus then one could say
that the nucleon momenta in the inital state are about $-{\bf q}_{AL}/$A
in this reference frame.  Absorption of a virtual photon of momentum ${\bf q}_{AL}$
by a ground state nucleon (the quasi-elastic process) would result in a final
state, where one nucleon has a momentum of about ${\bf q}_{AL}$(A-1)/A and
A-1 slower nucleons each have a momentum of about $-{\bf q}_{AL}/$A.

If one chooses to
minimize the sum of the center of mass kinetic energies of initial and final states, one is 
led to the Breit (B) frame.
In the Breit frame the target nucleus moves  with $-{\bf q}_B$/2 and the
nucleon momenta are thus about $-{\bf q}_B/$(2A).  According to the above picture  
the final state in the vicinity of the q.e. peak corresponds roughly to
one nucleon with momentum ${\bf q}_B$(2A-1)/(2A) and A-1 nucleons with
momenta of about $-{\bf q}_B$/(2A).
At fixed $q$ and $\omega$ values the anti--lab and Breit responses 
tend to the LAB response with increasing A.

As a fourth reference frame we introduce what we call the active
nucleon Breit frame (ANB).
In this frame  the target nucleus 
consisting of A nucleons has a momentum of $-{\rm A}{\bf q}_{ANB}/2$
so that the nucleons have momenta about -${\bf q}_{ANB}/2$ in the initial state.
The final state in the vicinity of the q.e. peak would correspond to an active 
nucleon with momentum of about ${\bf q}_{ANB}/2$ while the other nucleons continue
moving with the momenta about $-{\bf q}_{ANB}/2$. 
Thus within these approximations the 
maximum nucleon momentum is limited by $q_{ANB}/2\simeq q/2$ in the ANB frame, 
whereas in other reference frames nucleons with momenta up to $q$ are present. The momentum of 
the active nucleon is largest  
in the LAB frame so that one may expect that this reference frame is the least suitable 
within a n.r. approach. In particular, the relativistic correction related to
the kinetic energy is four times larger in the LAB frame than in the ANB frame. 
In the following we will calculate 
$R_L^{B}$, $R_L^{AL}$, and $R_L^{ANB}$ and then use (\ref{facs}) to give
the predicted $R_L$ from each of these. These indirectly calculated $R_L$
are then compared with $R_L$ as computed directly in the LAB frame at q=500, 600 and 700 MeV/c. 
By comparing our results to experimental data it should
become apparent if the ANB frame, for example, is superior to the LAB
frame.

The present 
calculation proceeds in the manner described in \cite{elot04}. There
we found  only a weak potential model dependence 
so that in the present calculation we choose the Argonne V18 (AV18) NN \cite{AV18}
plus Urbana IX (UrbIX) NNN \cite{Urb9} potentials. 
As in \cite{elot04} the n.r. charge operator is supplemented
with the first order relativistic corrections (Darwin--Foldy and spin--orbit terms).
However while in \cite{elot04} 
$q$ values up to 500 MeV/c were considered, in the present work we calculate the 
responses at $q=500$, 600 and 700 MeV/c. The inclusion of these higher q values
requires a larger set of basis states for convergence. For example, whereas 
the total angular momentum of the final states was limited to $J=21/2$ in \cite{elot04}
here we include states up to $J=31/2$.
As in \cite{elot04} we use the simple 
dipole fit for the proton electric form factor, but consider also the proton 
form factor fit from \cite{MMD}. For the neutron electric form factor we take 
the fit from \cite{Galster}.

Fig.~1 shows the $R_L$ results for the various frames together with 
experimental data at $q_{LAB}= 500,$ 600, and 700 MeV/c. It is readily seen 
that one obtains rather frame dependent results. One finds the following 
differences in peak positions and peak heights between the two extreme cases 
(ANB and LAB frame results): 6 MeV and 13\% (500 MeV/c), 11 MeV and 19\% (600
MeV/c), 20 MeV and 24\% (700 MeV/c). As anticipated of all four frames the LAB frame 
calculation leads to the worst result in comparison with experimental data. Let us
recall that these LAB results just represent the conventional n.r. 
calculation. On the other hand the ANB
frame leads to a good description of the data at $q=$500 and 600 MeV/c. This may be 
related to the fact that nucleons with only moderate momenta  are present in this
reference frame. Description 
of the data with the ANB frame would be even better if a contemporary proton form factor
in place of the dipole form factor is used. This will be demonstrated below.  
The above considerations demonstrate the frame dependence inherent in
a n.r. calculation of the longitudinal response at high q.
Clearly a proper relativistic calculation would remove this frame
dependence, but one can still ask whether there is a way to modify
the n.r. calculation such that the degree of frame dependence would
be reduced.

A clue is evident in the work of Arenh\"ovel and collaborators (see e.g. \cite{Fab78})
in deuteron electrodisintegration, where the relative momentum of
outgoing nucleons is determined in a relativistically  correct
way and the energy that is used as input to the n.r. calculation
is obtained from that momentum by the usual n.r. relation.
In general in a two--body problem one may either determine 
the kinetic energy in a relativistically correct way
and solve the n.r. Schr\"odinger 
equation with it  or determine the relative momentum 
$p_{12}$ in a relativistically correct way
and solve the Schr\"odinger equation for the "fake" kinetic energy $E_{12}=
p^2_{12}/2\mu_{12}$, where $\mu_{12}$ is the reduced mass of the two particles.
The reason why the latter procedure is chosen in the case of deuteron
electrodisintegration is because the construction of NN potential models proceeds that way.

If one is mainly interested in the region of the quasielastic peak then one
can adopt an analogous procedure based on a two--body model for the quasi-elastic
process.  That is, the final state is assumed to consist of a knocked--out nucleon
and an (A-1) particle residual system remaining in its lower energy state. 
We stress here that the two--body model is adopted only for determining 
the kinematical input of a calculation where the full three--body dynamics
is properly taken into account.

The momenta of the knocked--out nucleon and that of the residual nucleus
are denoted by ${\bf p}_N^{fr}$ and ${\bf p}_X^{fr}$, respectively. 
Then the relative and center of mass momenta will be given by
${\bf p}^{fr}=\mu({\bf p}_N^{fr}/M-{\bf p}_X^{fr}/M_X)$ and
${\bf P}_f^{fr}={\bf p}_N^{fr}+{\bf p}_X^{fr}$, where $M_X$ is the mass
of the residual nucleus and $\mu$ is the $N-X$ reduced mass. (Note that
${\bf p}^{fr}$ depends on the reference frame in the relativistic case). 
The value of ${\bf p}^{fr}$ can be obtained from the following relativistically 
correct kinematical relation
\begin{equation}\label{omegafr}
\omega^{fr}=E_f^{fr}-E_i^{fr}
\end{equation}
where
\begin{equation}
E_f^{fr}\ =\ \sqrt{M^2+[{\bf p}^{fr}+(\mu/M_X){\bf P}_f^{fr}]^2}+
\sqrt{M_X^2+[{\bf p}^{fr}-(\mu/M){\bf P}_f^{fr}]^2}
\end{equation}

Then, in accordance with the preceding discussion on the two--body system, the final state relative energy 
to use in the n.r. calculation is taken to be
\begin{equation}\label{enr}
e_f^{fr}=(p^{fr})^2/(2\mu)\,.
\end{equation}
Here one has to notice that in order to solve Eq. (\ref{omegafr}) for $p^{fr}$ one needs to know its direction.
For the class of reference frames we consider the momentum ${\bf P}_f^{fr}$
is directed along ${\bf q}$. Again, since we are mainly interested 
in the region of the quasielastic peak we can safely assume that 
${\bf p}^{fr}$ is also directed along ${\bf q}$.
(Indeed, e.g. ${\bf p}^{LAB}\simeq(\mu/M){\bf q}$.)

Proceeding in the way described above is formally equivalent to 
replace $(E_f^{fr}-E_i^{fr})$ in the delta function of Eq. (\ref{rfr}) by a function
$F(e_f^{fr})=(E_f^{fr}(e_f^{fr})-E_i^{fr})$. Therefore
\begin{equation}
\delta\left(E_f^{fr}-E_i^{fr}-\omega^{fr})\right)\,
=\left(\frac{\partial F^{fr}}{\partial e_f^{fr}}\right)^{-1}
\delta\left(e_f^{fr}-e_f^{rel}(q^{fr},\omega^{fr})\right),
\end{equation}
with 
\begin{equation}
\left(\frac{\partial F^{fr}}{\partial e_f^{fr}}\right)^{-1}=\frac{p^{fr}}{\mu}
\left(\frac{\partial E_f}{\partial p^{fr}}\right)^{-1}\,.
\end{equation} 
This leads to
\begin{equation}
R_L^{fr}(q^{fr},\omega^{fr})=\\
\frac{p}{\mu}
\left(\frac{\partial E_f}{\partial p}\right)^{-1}\,\sum\!\!\!\!\!\!\!\int df
\left|\langle\psi_i|\sum_j\rho_j(q^{fr},\omega^{fr})|\psi_f\rangle\right|^2
\delta\left(e_f^{fr}-e_f^{rel}(q^{fr},\omega^{fr})\right)\,.
\end{equation} 
In order to calculate this quantity a new calculation is not required. We have  obtained it
via interpolation with respect to the momentum transfer of the n.r. response.

This procedure should reduce the frame dependence of $R_L(q,\omega)$ considerably.
This is evident in the free case, i.e. when there is no interaction between the fast nucleon 
and the residual system. In this case the n.r. and relativistic final states
would contain the relative
motion plane wave with the same momentum ${\bf p}$,
resulting in no frame dependence of the matrix elements due to a difference in relative
motion. This situation can be
only slightly changed by the nuclear force. 

In Fig.~2 we 
show the various $R_L$ results in comparison with experimental data and
in fact we find an enormous reduction of the frame dependence. For the peak
positions we even have an essentially frame independent result and also the 
differences of the peak heights are much reduced, namely to maximally 4, 6 and 
9\% at $q$=500, 600 and 700 MeV/c, respectively. It is evident that
there is a good agreement between theory and experiment for the position of
the quasielastic peak at all three momentum transfers. Concerning the peak
heights one finds a relatively good agreement at $q$=500 and 600 MeV/c,
while at 700 MeV/c the theoretical peak height overestimates the experimental one between
about 20 and 30\%.

It is interesting to check which of the frame dependent results of Fig.~1
reproduces best the frame independent peak positions of Fig.~2. It turns out that
this is the ANB frame. Also the peak heights of the ANB curves in Fig.~1 and  Fig.~2
are not much different: 4\% 
($q=500$ MeV/c), 5\% ($q=600$ MeV/c), 6\% ($q$=700 MeV/c). This is not surprising since 
the ANB frame is the only frame where the nucleon has equal initial and final 
energies (in fact the initial
nucleon momentum is about $-{\bf q}_{ANB}/2$  and its final momentum is ${\bf q}_{ANB}/2$).
Thus the quasi-elastic peak occurs at $\omega_{ANB}=0$, independent of whether 
relativistic or n.r. kinematics are employed.
Note that in the A=2 case the ANB frame
coincides with the anti--lab frame which is often chosen for
the deuteron electrodisintegration. 

A comparison of Figs.~1 and ~2 shows that the n.r. ANB frame
calculations not only agree with the relativistic two-body kinematics
calculations at the peak but also in the tails. This is illustrated
in another way in Fig.~3 where the n.r. ANB  frame results are
shown together with the relativistic kinematics Breit frame 
results.  The choice of the Breit frame was motivated by the
deuteron electrodisintegration work of \cite{BA3} where it was
shown that boost corrections are minimal for this frame.

Apart from
theoretical uncertainties of the quasielastic $R_L$ response due to 
frame dependence,
probably the greatest remaining theoretical 
uncertainty is due to the proton electric form factor. As an
illustration we show in Fig.~4 the Breit frame results with relativistic 
two--body kinematics using the two above mentioned different proton electric 
form factors (dipole fit and fit from \cite{MMD}). In comparison to the
dipole fit the fit of \cite{MMD} reduces the peak height by about 4, 6 and
7\% at $q=500$, 600 and 700 MeV/c leading to an improved agreement with
experiment at the lower two $q$ values and reducing the discrepancy at 
$q=700$ MeV/c to about 15\%. On the other hand,
the rather large experimental uncertainties preclude making
definitive conclusions.

We summarize our results as follows. We have shown that the usual 
n.r. calculation of the longitudinal inclusive $(e,e')$ response
leads to rather frame dependent results at intermediate momentum
transfers of $q=500-700$ MeV/c. The frame dependence is drastically reduced
if one assumes a two--body break up model with relativistic kinematics to determine
the input to the n.r. dynamics calculation. One obtains
a nearly frame independent peak position and much smaller deviations for 
the peak heights. Within n.r.
kinematics, of the considered reference frames the ANB
frame turns out to be the best, leading to results almost identical 
to those obtained  with the suggested two--body break up model. In comparison 
with experimental data we find good
agreements for the positions of the quasielastic peak and also good
agreements of the peak heights at $q=500$ and 600 MeV/c, while at $q=700$
MeV there is a discrepancy between about 15 and 25\%.

We thank J. Jourdan for providing us with the data of Ref.~\cite{World}.
V.D.E. acknowledges the support from the RFBR, grant 05-02-17541, and the Russian Ministry
of Industry and Science, grant NS--1885.2003.2.  E.L.T.
acknowledges support from the National Science and Engineering Research Council of Canada.
This work was supported by the grant COFIN03 of the Italian Ministery of University and Research.

\section{Bibliograhy}

\vfill\eject

\centerline{\bf FIGURE CAPTIONS}

\bigskip\bigskip

\noindent FIG.1: Frame dependence of the $^3$He longitudinal response function at three
different momentum transfers $q$ (notation of curves in upper panel); 
also shown experimental data from Refs. \cite{Saclay} (squares), \cite{Bates}
(triangles), \cite{World} (circles).

\bigskip

\noindent FIG.2: As Fig.~1, but considering two-body relativistic kinematics for the
final state energy as discussed in the text.

\bigskip

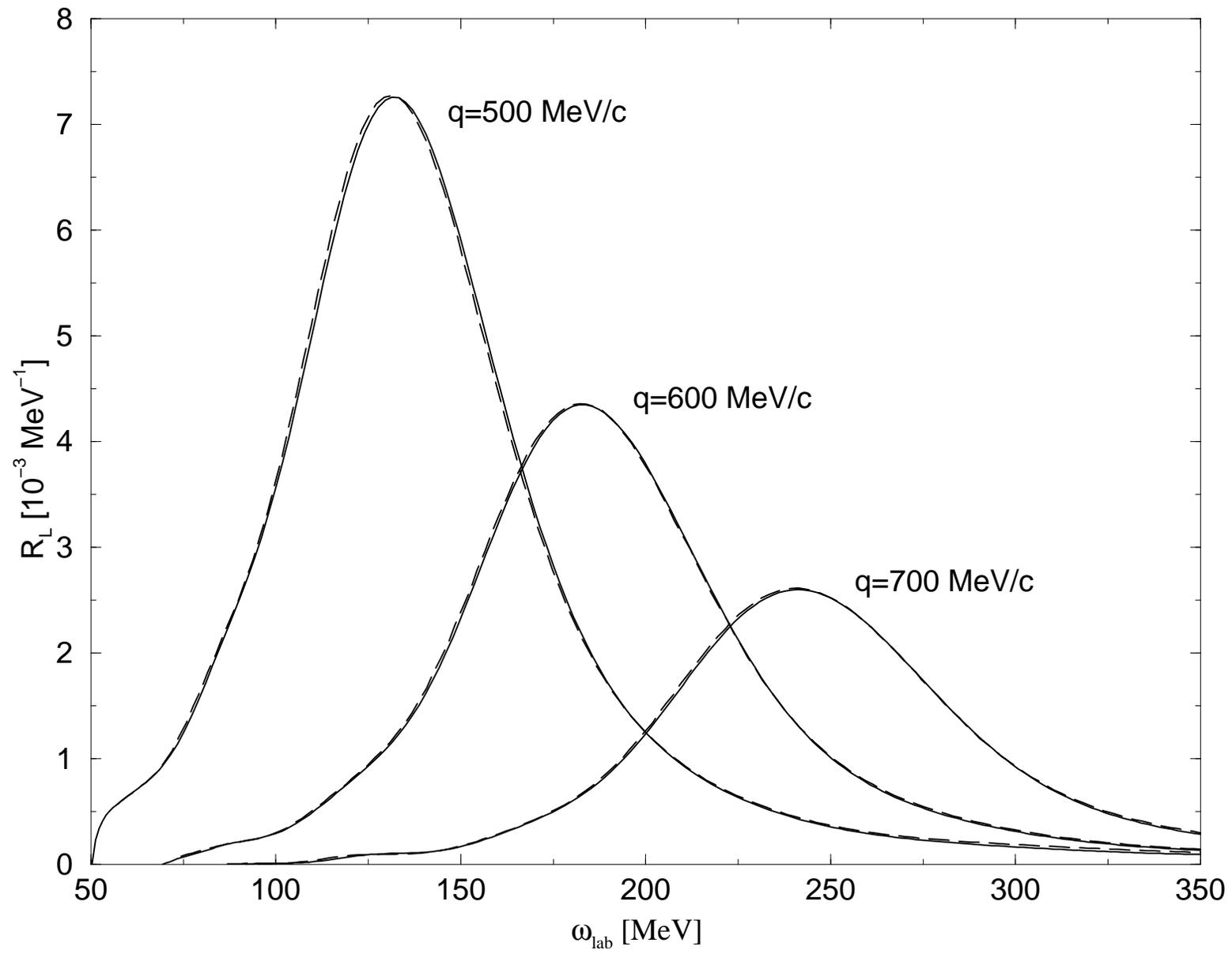
\noindent FIG.3: $R_L$ of ANB frame calculations without consideration of two-body relativistic kinematics
(long dashed curves) in comparison to $R_L$ of B frame calculations with consideration of
two-body kinematics (full curves).

\bigskip

\noindent FIG.4: $R_L$ of B frame calculations with consideration of two-body relativistic kinematics 
using different proton electric form factors: dipole fit (full curves), fit from 
\cite{MMD} (long dashed curves); notation of experimental data as in Fig.~1.

\bigskip

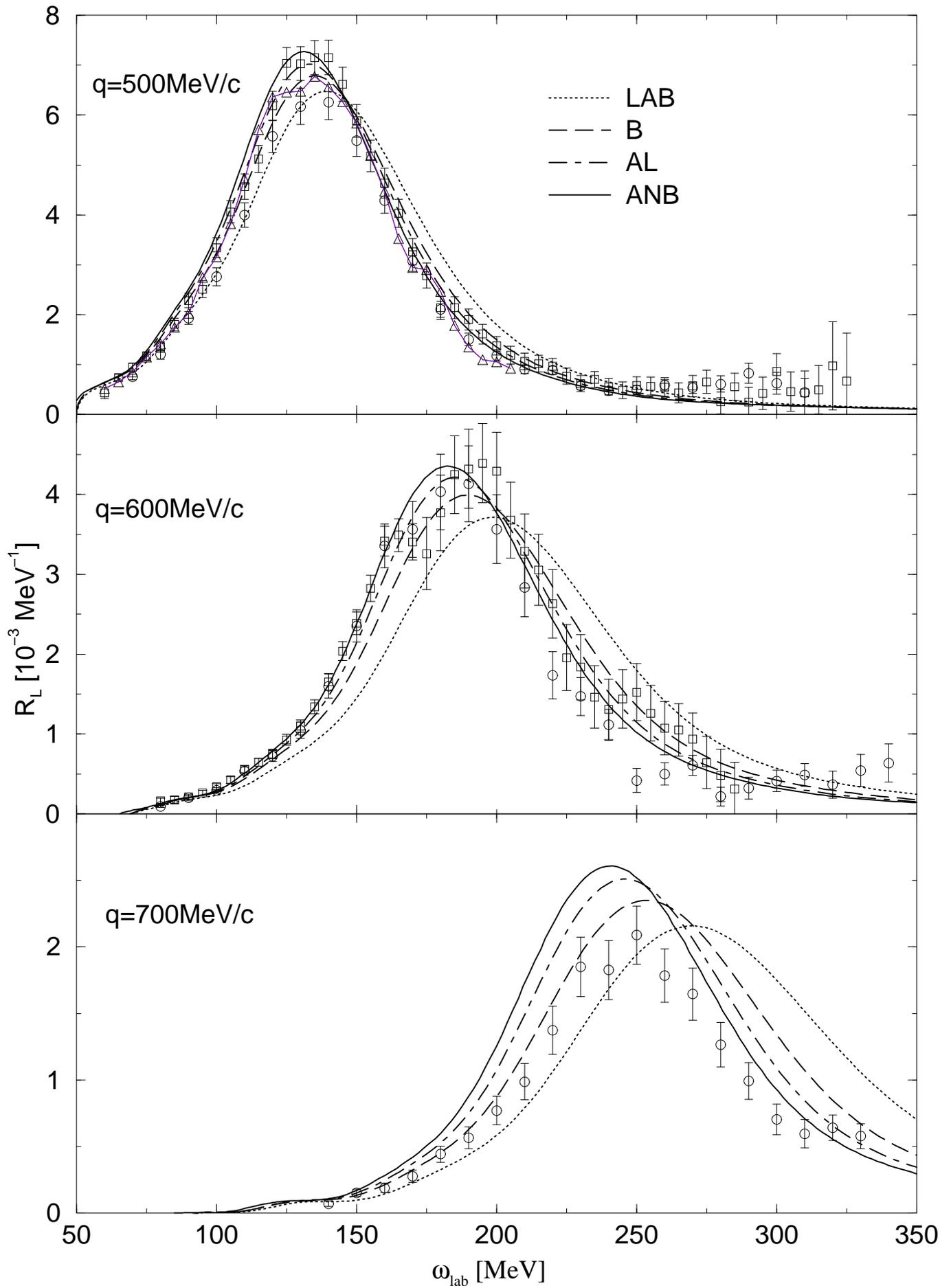
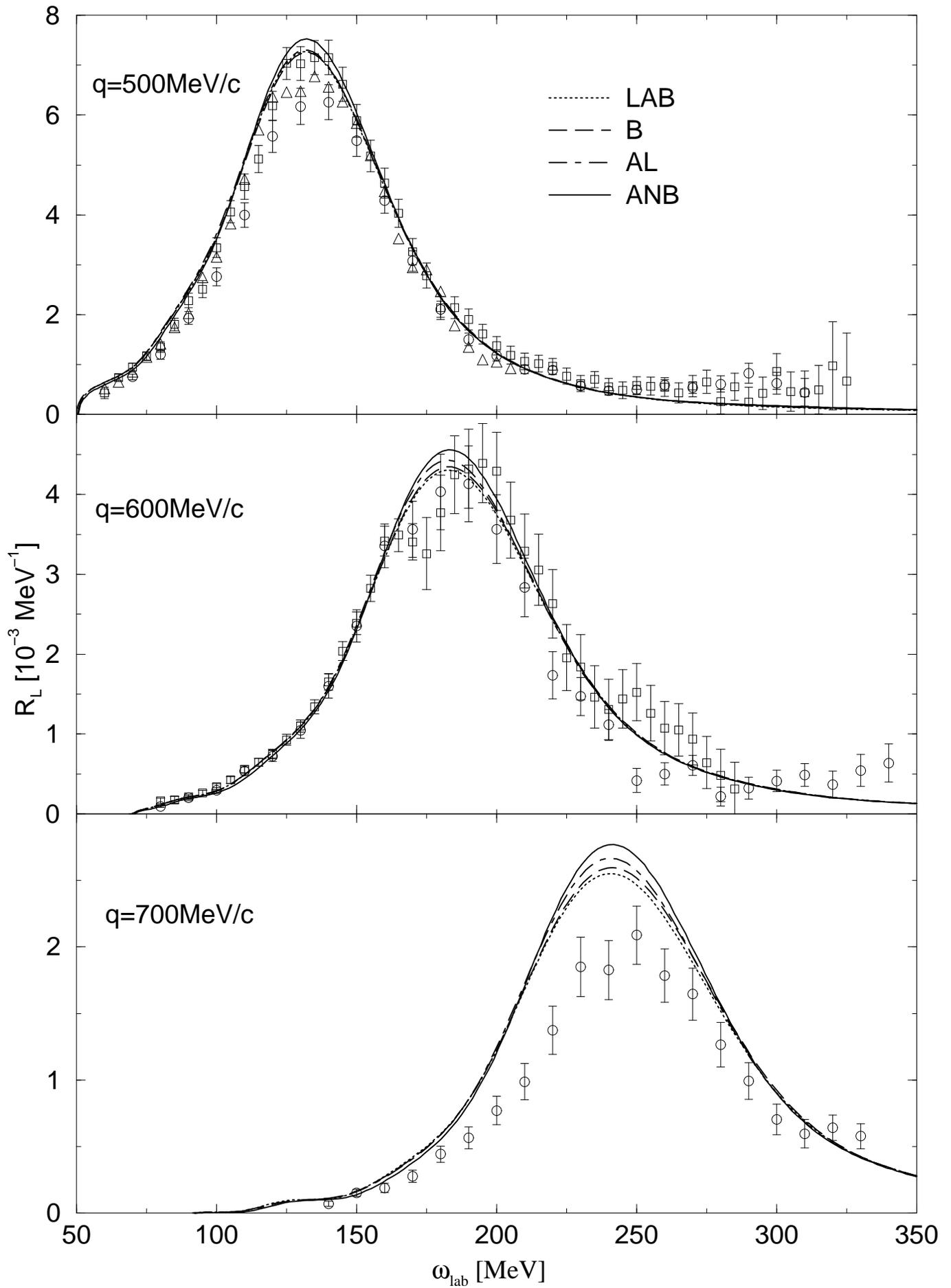
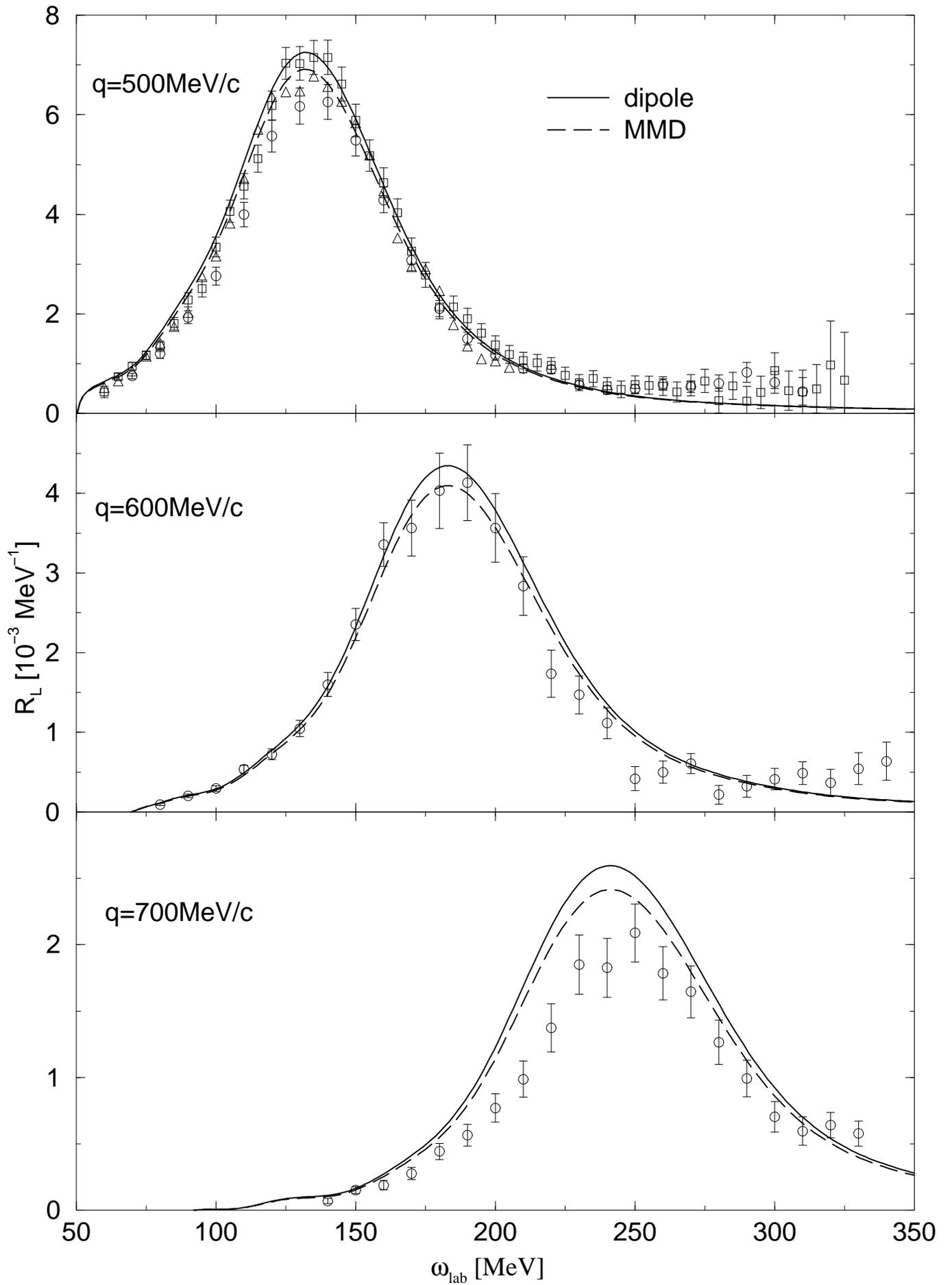
\end{document}